\documentclass[submission,copyright,creativecommons]{eptcs}

\providecommand{\event}{NCL'26} 

\usepackage{iftex}
\ifpdf
  \usepackage{underscore}         
  \usepackage[T1]{fontenc}        
\else
  \usepackage{breakurl}           
\fi
\usepackage[normalem]{ulem} 

\usepackage{amssymb,mathtools,amsthm}
\usepackage{stmaryrd}
\usepackage{enumitem}
\usepackage[dvipsnames,svgnames,x11names]{xcolor}
\usepackage[disable]{todonotes}

\usepackage{tikz}
\usetikzlibrary{positioning}

\theoremstyle{plain}
\newtheorem{theorem}{Theorem}[section]  
\newtheorem{lemma}[theorem]{Lemma}    

\newtheorem{proposition}[theorem]{Proposition}

\theoremstyle{definition}
\newtheorem{definition}[theorem]{Definition}

\theoremstyle{remark}

\newcommand{\jt}{\mathbin{\rightarrow}} 
\newcommand{\n}{\neg}
\newcommand{\et}{\mathbin{\wedge}}

\newcommand{\Fm}{\mathrm{Fm}}
\newcommand{\Prop}{\mathrm{Prop}}

\newcommand{\CLuN}{\mathsf{CLuN}}

\newcommand{\Mn}{\mathcal M_{n}}

\newcommand{\Pow}{\mathcal P}
\newcommand{\set}[1]{\{#1\}}
\newcommand{\interp}[1]{\llbracket #1\rrbracket}
\newcommand{\nt}{\interp{\n}}
\newcommand{\wtwedge}{\interp{\wedge}}
\newcommand{\wtvee}{\interp{\vee}}
\newcommand{\wtto}{\interp{\to}}
\newcommand{\nmodelsM}{\not\models_{\Mn}}
\newcommand{\modelsM}{\models_{\Mn}}

\newcommand{\vdashn}{\vdash_{\mathbf{n}}}
\newcommand{\nvdashn}{\nvdash_{\mathbf{n}}}

\title{Uniform Non-Deterministic Semantics for Ciuciura's \texorpdfstring{$\mathsf{mbC}^{n}$}{mbC^n}-Hierarchy}

\author{Ela Drozdowska
\institute{The John Paul II Catholic University of Lublin}
\and Pawel Pawlowski\thanks{Pawel Pawlowski acknowledges funding from the Research Foundation Flanders (FWO) under grant number 1255724N.}
\institute{CLPS, Ghent University\\ Belgium}
}

\def\titlerunning{Uniform NDS for Ciuciura's mbC\textsuperscript{n}-Hierarchy}
\def\authorrunning{E. Drozdowska \& P. Pawlowski}
\def\copyrightholders{Ela Drozdowska and Pawel Pawlowski}

\begin{document}
\maketitle

\begin{abstract}
We introduce a uniform non-deterministic semantics for the hierarchy of paraconsistent systems $\mathbf{mbC}^{n}$ introduced by Ciuciura. The proposed construction generalizes Omori’s three-valued semantics for $\mathbf{mbC}^{1}$ to all systems of the hierarchy. We prove soundness and completeness of the semantics with respect to the corresponding axiomatizations and show that the consequence relations of the systems $\mathbf{mbC}^{n}$ cannot be characterized by finite deterministic matrices.
\end{abstract}


\section{Introduction}

From a bird's-eye perspective, a logic is \emph{paraconsistent} iff
its consequence relation is non-explosive, i.e.,  a contradiction does not entail everything. In the classical setting, \textit{ex contradictione quodlibet} (the law of logical explosion) lets you derive 
any conclusion $\psi$ from the set of inconsistent premises. Thus a single contradiction is enough to trivialize a~whole system and render it useless in the contexts where premises may be contradictory. Paraconsistent logics were created to deal with exactly such contexts.\footnote{Historically, Vasiliev is often regarded as an important forerunner of paraconsistent logic. However, the first formal paraconsistent system was developed by  the Polish logician Stanisław Jaśkowski \cite{Jaskowski1948} in the form of a discussive logic.} They block that problematic inference, requiring that for  \emph{some} $\varphi, \psi$, the entailment $\{\varphi, \neg \varphi\} \vdash \psi $ fails.  This in turn allows one to reason in spite of the presence of contradictions without collapse.\footnote{According to \cite{DaCostaBeziuBueno1995}, Francisco Mir\'o Quesada Cantuarias considered several possible labels, including \emph{metaconsistent}, \emph{ultraconsistent}, and \emph{paraconsistent} for this class of logics. Ultimately, the term paraconsistent was adopted and  was first officially proposed at the Third Latin American Symposium on Mathematical Logic in 1976.}

One of the most common ways of introducing paraconsistency\footnote{Paraconsistent logic now comprises a broad family of various approaches to non-explosive consequence relations \cite{DaCosta1974,CarnielliConiglioMarcos2007,RescherManor1970,SchotchBrownJennings2009,UrbaniakSinilo2014}.} is through the introduction of gluts, i.e., states in which a formula and its negation are both true. A paradigmatic example of this strategy is the logic developed by Diderik Batens\footnote{It was known as PI.} \cite{Batens1980}, \emph{classical logic with gluts} $\mathbf{CLuN}$.\footnote{$\mathbf{CLuN}$ belongs to a small family of logics built on the same positive classical base and differing only in the strength of their negation. Alongside it sit two others: $\mathbf{CLoN}$, whose negation is very weak and connects with super weak modal systems, and $\mathbf{CLaN}$, the paracomplete member, whose negation admits gaps rather than gluts.} This logic validates the positive classical principles for $\wedge,\vee,\to$ and the principle of excluded middle $\varphi\vee\neg\varphi$. It does not validate unrestricted explosion.

$\mathbf{CLuN}$ has served as the starting point for two major paradigms in paraconsistent logic: adaptive logics and logics of formal inconsistency (LFI). In this paper, we focus on the latter. We examine a variant of the LFIs, formulated in the classical language: the systems $\mathbf{mbC}^{n}$, introduced by the Polish logician Janusz Ciuciura\footnote{The notation follows Ciuciura's original papers and Omori's subsequent work \cite{Omori2019}. In his later monograph \cite{Ciuciura2018}, however, Ciuciura remarks that the notation $\mathbf{mbC}^{n}$ may be misleading, as it suggests a connection with the logic $\mathbf{mbC}$ of Carnielli and Marcos, in particular the equivalence of $\mathbf{mbC}$ and $\mathbf{mbC}^{1}$. To avoid this ambiguity, he adopted the notation $\mathbf{B}^{n}$. Since the present paper builds directly on the work of Omori, we retain the original notation $\mathbf{mbC}^{n}$.} \cite{Ciuciura2014}. Like other LFIs, the systems in Cuciura's hierarchy are paraconsistent. However, unlike most LFIs, they are formulated in the classical propositional language without the additional consistency operator. Moreover, what sets these logics apart is the fact that they allow a restricted  form of controlled explosion. This means that under certain additional assumption, the systems validate a variant of the explosion principle. The hierarchy is indexed by a parameter $n$, which determines how explosion is recovered through iterated negations. As $n$ increases, longer chains of iterated negations are required before explosion can be recovered.

In this paper, we continue the work of Omori \cite{Omori2019}, who introduced a three-valued non-deterministic semantics for the first level of Ciuciura's hierarchy. We extend this approach by providing a uniform non-deterministic semantics for the full hierarchy, which we call the $\mathbf{mbC}^{n}$-matrices. Let us continue with a short introduction of what non-deterministic semantics is.

Non-deterministic semantics generalizes classical matrix semantics by allowing the interpretation of a connective to return a non-empty set of values, rather than a single value. A valuation may then pick any value from that set, and the classical matrices are recovered as the special case in which every such set is a singleton. This is exactly what lets one handle a connective that does not behave classically, such as the negation of the $\mathbf{mbC}^{n}$-systems.

Non-deterministic semantics has been studied most extensively in the setting of modality. The first attempt to link non-deterministic semantics with modal operators was made in \cite{Kearns1978,Kearns1981,Ivlev1991}. Subsequently, the work of Coniglio and Omori \cite{Coniglio2015,Omori2016} established a strong paradigm of non-deterministic semantics for propositional modality \cite{Kearns1981,Ivlev1988,Coniglio2015,Coniglio2019,Coniglio2019b,Omori2020,Omori2021,Pawlowski2021,Pawlowski2022,Pawlowski2022a,Graetz2022,Omori2024,Pawlowski2024,ConiglioPawlowskiSkurt2025,Pawlowski2024a,Coniglio2024,Leme2025,Pawlowski2026Boolean,vaz2025swapkripkemodelsdeontic}. It has also been applied to logics of formal inconsistency \cite{Avron2005b,Avron2007d,Avron2009,Omori2019}, informal provability \cite{Pawlowski2018,Pawlowski2022}, a broader class of paraconsistent logics \cite{Avron2017,Avron2005}, and even quantum logic \cite{Jorge2020}. Omori showed that Ciuciura's system $\mathbf{mbC}^1$ admits a three-valued non-deterministic semantics \cite{Omori2019}.

The remainder of the paper is organized as follows. Section~2 contains the necessary preliminaries. In Section~3, we recall Ciuciura's hierarchy and its original semantics. We also show that $\mathbf{mbC}^n$ is not characterized by any finite deterministic logical matrix with unrestricted valuations. In Section~4, we define $\mathbf{mbC}^{n}$-matrices and prove soundness and completeness of the resulting semantics. Finally, Section~5 contains some concluding remarks.

\section{Technical preliminaries}
Throughout the paper, we use \(\Prop\) for a countable set of propositional variables and \(\Fm\) for the set of all formulas generated from \(\Prop\) by the connectives \(\n,\vee,\et,\jt\) of the classical propositional language. Lowercase Greek letters \(\varphi,\psi,\dots\) are used to denote formulas, while uppercase Greek letters \(\Gamma,\Pi,\dots\) denote sets of formulas. Furthermore, we use \(\n^{n}\varphi\) as an abbreviation for \(\underbrace{\n\,\ldots\,\n}_{n\ \mathrm{times}}\varphi\).

On the semantic side, we work with non-deterministic semantics. This generalizes classical matrix semantics by allowing the interpretation of a connective to return a non-empty set of values, rather than a single value. Formally, a non-deterministic matrix, or Nmatrix, for this language is a triple \(\mathcal M=\langle V,D,O\rangle\), where \(V\) is a non-empty set of values, \(D\subseteq V\) is the non-empty set of designated values, and \(O\) assigns non-empty multioperations to the logical connectives:
\begin{enumerate}

\item

\(\interp{\n}:V\to\Pow(V)\setminus\{\varnothing\}\).

\item \(\interp{\wedge},\interp{\vee},\interp{\to}:V^2\to\Pow(V)\setminus\{\varnothing\}\).

\end{enumerate}
\noindent A valuation in such a matrix is a function \(v:\Fm\to V\) such that \(v(\n\varphi)\in\interp{\n}(v(\varphi))\) and furthermore, \[
v(\varphi\circledast\psi)\in\interp{\circledast}(v(\varphi),v(\psi))
\]
for every \(\circledast\in\{\wedge,\vee,\to\}\). A formula $\varphi$ is satisfied by a valuation $v$, written \(v\models_{\mathcal M}\varphi\), iff \(v(\varphi)\in D\). The induced consequence relation is defined, as usual, by preservation of designatedness over all valuations of the Nmatrix.

\section{The hierarchy}

This section introduces the hierarchy. Since all the technical ingredients are now set up, let us proceed with the formal definition of systems $\mathbf{mbC}^{n}$. Axiomatically, they can be seen as extensions of the paraconsistent logic $\mathbf{CLuN}$ obtained by adding the corresponding Ciuciura's axioms. First, we home in on the base system $\mathbf{CLuN}$.

Axiomatically, $\mathbf{CLuN}$ consists of the positive fragment of classical propositional logic together with the law of excluded middle:
\begin{center}
\begin{minipage}[t]{0.48\textwidth}
\small
\begin{align*}
(\mathrm{A}_1)&\quad \varphi\to(\psi\to\varphi),\\[.35ex]
(\mathrm{A}_2)&\quad \begin{aligned}[t]
(\varphi\to(\psi\to\chi))\to{}&((\varphi\to\psi)\to{}\\
& (\varphi\to\chi)),
\end{aligned}\\[.35ex]
(\mathrm{A}_3)&\quad \varphi\to(\psi\to(\varphi\wedge\psi)),\\[.35ex]
(\mathrm{A}_4)&\quad (\varphi\wedge\psi)\to\varphi,\\[.35ex]
(\mathrm{A}_5)&\quad (\varphi\wedge\psi)\to\psi.
\end{align*}
\end{minipage}\hfill
\begin{minipage}[t]{0.48\textwidth}
\small
\begin{align*}
(\mathrm{A}_6)&\quad \varphi\to(\varphi\vee\psi),\\[.35ex]
(\mathrm{A}_7)&\quad \psi\to(\varphi\vee\psi),\\[.35ex]
(\mathrm{A}_8)&\quad \begin{aligned}[t]
(\varphi\to\chi)\to{}&((\psi\to\chi)\to{}\\
&((\varphi\vee\psi)\to\chi)),
\end{aligned}\\[.35ex]
(\mathrm{A}_9)&\quad \varphi\vee(\varphi\to\psi),\\[.35ex]
(\mathrm{A}_{10})&\quad \varphi\vee\n\varphi.
\end{align*}
\end{minipage}
\end{center}
The only inference rule is modus ponens:
\[
\frac{\varphi\qquad \varphi\to\psi}{\psi}\; (\mathrm{MP}).
\]

For a set of formulas \(\Gamma\) and a formula \(\varphi\), we write \(\Gamma \vdash_{\mathbf{CLuN}} \varphi\) if \(\varphi\) is derivable from \(\Gamma\) in $\mathbf{CLuN}$.

Note that the Hilbert basis used here differs from Ciuciura's original basis \cite{Ciuciura2014} in the ninth axiom. We use

$$(A_9)\qquad \varphi\vee(\varphi\to\psi),$$

\noindent whereas Ciuciura uses Peirce's law

$$(P)\qquad ((\varphi\to\psi)\to\varphi)\to\varphi.$$

However, the two axiomatizations are equivalent\footnote{The two positive bases are deductively equivalent. From $(\mathrm A_9)$ one derives Peirce's law $((\varphi\jt\psi)\jt\varphi)\jt\varphi$: writing $H:=(\varphi\jt\psi)\jt\varphi$, an instance of $(\mathrm A_8)$ (with $\chi:=\varphi$) together with $\vdash\varphi\jt\varphi$ gives $H\jt((\varphi\vee(\varphi\jt\psi))\jt\varphi)$; since $\varphi\vee(\varphi\jt\psi)$ is a~theorem by $(\mathrm A_9)$, $(\mathrm A_1)$ gives $H\jt(\varphi\vee(\varphi\jt\psi))$, and chaining the two implications via $(\mathrm A_2)$ yields $H\jt\varphi$. Conversely, Peirce's law gives $(\mathrm A_9)$: put
$\chi:=\varphi\vee(\varphi\jt\psi)$ and assume $\chi\jt\varphi$. By $(\mathrm A_7)$,
$(\varphi\jt\psi)\jt\chi$, so chaining gives $(\varphi\jt\psi)\jt\varphi$; Peirce's law
(instance with second argument $\psi$) then yields $\varphi$, and $(\mathrm A_6)$ gives
$\chi$. Discharging the assumption gives $\vdash(\chi\jt\varphi)\jt\chi$; instantiating
Peirce's law with $\chi$ in place of $\varphi$ gives $((\chi\jt\varphi)\jt\chi)\jt\chi$,
whence $\vdash\chi$, i.e. $(\mathrm A_9)$.}.

We now extend $\mathbf{CLuN}$ with Ciuciura's axioms, that are parametrized by a natural number $n$. Intuitively, the parameter \(n\) determines how many successive negations may occur before arbitrary conclusions become derivable. 
\begin{definition}[Ciuciura's hierarchy]\label{def:hierarchy}
We start by introducing the axioms regulating the hierarchy. The \((n+1)\)-st Ciuciura axiom schema is:
\[
(\mathtt{Ciuciura}_{n+1})\qquad
\mathtt{Ciuciura}_{n+1}(\varphi,\psi)
:=
\varphi\to(\n\varphi\to(\n^{2}\varphi\to\cdots(\n^{n+1}\varphi\to\psi)\cdots)).
\]
The system $\mathbf{mbC}^{n}$ is obtained by extending $\mathbf{CLuN}$ with the axiom schema \(\mathtt{Ciuciura}_{n+1}\) for all formulas of the language. On this reading $\mathbf{mbC}^{0}:=\mathbf{CLuN}+ \mathtt{Ciuciura}_{1}=\mathbf{CPL}$. Following Ciuciura's convention \cite{Ciuciura2018}, we also stipulate that $\mathbf{mbC}^{\omega}=\mathbf{CLuN}$. Furthermore, for a set of formulas \(\Gamma\) and a formula \(\varphi\), we write \(\Gamma \vdashn \varphi\) if \(\varphi\) is derivable from \(\Gamma\) in $\mathbf{mbC}^{n}$.

\end{definition}

Some auxiliary comments are in order. First, recall that for \(n=0\), Ciuciura's axiom takes the form of the classical law of explosion:
\[
(\mathtt{Ciuciura}_{1})\qquad
\mathtt{Ciuciura}_{1}(\varphi,\psi)=
\varphi\to(\neg\varphi\to\psi).
\]

If \(n=1\), Ciuciura's axiom takes the form
\[
(\mathtt{Ciuciura}_{2})\qquad
\mathtt{Ciuciura}_{2}(\varphi,\psi)=
\varphi\to(\n\varphi\to(\n\n\varphi\to\psi)).
\]

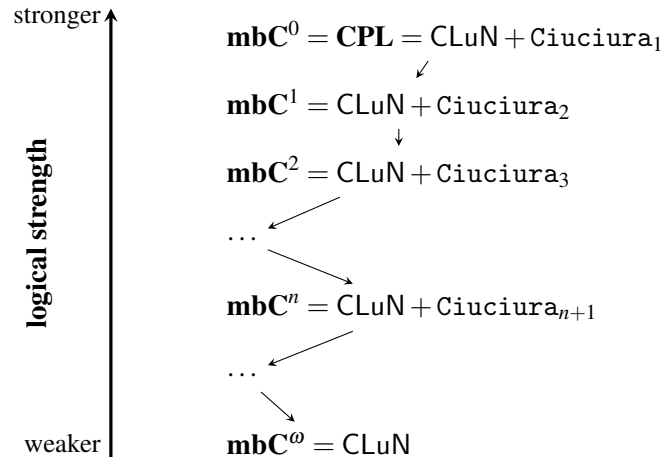
\begin{figure}[ht]
\centering
\begin{tikzpicture}[x=1cm,y=0.9cm,>=stealth]

  \draw[->, very thick] (-1.4,-0.2) -- (-1.4,6.4);
  \node[rotate=90] at (-2.35,3.1) {\textbf{logical strength}};
  \node[left] at (-1.4,6.25) {\small stronger};
  \node[left] at (-1.4,0.00) {\small weaker};

  \node[anchor=west] (mbc0) at (0,6)
    {\(\mathbf{mbC}^{0}=\mathbf{CPL}=\CLuN+\mathtt{Ciuciura}_{1}\)};

  \node[anchor=west] (mbc1) at (0,5)
    {\(\mathbf{mbC}^{1}=\CLuN+\mathtt{Ciuciura}_{2}\)};

  \node[anchor=west] (mbc2) at (0,4)
    {\(\mathbf{mbC}^{2}=\CLuN+\mathtt{Ciuciura}_{3}\)};

  \node[anchor=west] (dots1) at (0,3)
    {\(\cdots\)};

  \node[anchor=west] (mbcn) at (0,2)
    {\(\mathbf{mbC}^{n}=\CLuN+\mathtt{Ciuciura}_{n+1}\)};

  \node[anchor=west] (dots2) at (0,1)
    {\(\cdots\)};

  \node[anchor=west] (mbcomega) at (0,0)
    {\(\mathbf{mbC}^{\omega}=\CLuN\)};

  \draw[->] (mbc0) -- (mbc1);
  \draw[->] (mbc1) -- (mbc2);
  \draw[->] (mbc2) -- (dots1);
  \draw[->] (dots1) -- (mbcn);
  \draw[->] (mbcn) -- (dots2);
  \draw[->] (dots2) -- (mbcomega);

\end{tikzpicture}
\caption{The \(\mathbf{mbC}^{n}\)-hierarchy ordered by logical strength.}
\label{fig:mbc-hierarchy}
\end{figure}

So, the above is the core idea of the entire hierarchy. Intuitively, it says that  in order to obtain a~controlled version of the explosion principle, we need not only $\varphi,\n \varphi$ but also the appropriate iterated negation, depending exactly on the parameter of the system in the hierarchy. Furthermore, the hierarchy forms a linear order with respect to logical strength understood in terms of derivability.\footnote{ Derivability strength is understood as inclusion between sets of theorems: 
\(\mathsf L_1\) is stronger than \(\mathsf L_2\) iff every theorem of 
\(\mathsf L_2\) is a theorem of \(\mathsf L_1\). One may also think about these systems in terms of the corresponding consequence relation and the hierarchy would be the same.} Consequently, $\mathbf{mbC}^{0}$ is the strongest system, while the systems further up in the hierarchy are subsystems of the preceding systems. This is nicely summarized by the diagram in Figure~\ref{fig:mbc-hierarchy}.

We continue with a minor observation showing that the hierarchy is well defined, since every system proves the characteristic axioms of every weaker system.
\begin{proposition}\label{hierarchy-inclusion}
For every \(n\geq 1\) and every \(m\geq n+1\),
\(\vdash_{\mathbf{mbC}^{n}}\mathtt{Ciuciura}_{m}(\varphi,\psi)\).
\end{proposition}

\begin{proof}
If \(m=n+1\), the claim is immediate, since
\(\mathtt{Ciuciura}_{n+1}\) is the characteristic axiom schema of
\(\mathbf{mbC}^{n}\). Suppose \(m>n+1\), and put
\(\chi:=\neg^{n+2}\varphi\to(\neg^{n+3}\varphi\to\cdots
(\neg^{m}\varphi\to\psi)\cdots)\). Since
\(\mathtt{Ciuciura}_{n+1}\) is an axiom schema of
\(\mathbf{mbC}^{n}\), we may instantiate it with this \(\chi\) in place
of \(\psi\). This yields
\(\varphi\to(\neg\varphi\to\cdots
(\neg^{n+1}\varphi\to\chi)\cdots)\), which, by the definition of
\(\chi\), is exactly \(\mathtt{Ciuciura}_{m}(\varphi,\psi)\).
Hence \(\vdash_{\mathbf{mbC}^{n}}\mathtt{Ciuciura}_{m}(\varphi,\psi)\).
\end{proof}

For example, in \(\mathbf{mbC}^{1}\) the schema
\(\varphi\to(\neg\varphi\to(\neg^{2}\varphi\to\chi))\) gives
\(\mathtt{Ciuciura}_{3}(\varphi,\psi)\) by taking
\(\chi:=\neg^{3}\varphi\to\psi\). Taking instead
\(\chi:=\neg^{3}\varphi\to(\neg^{4}\varphi\to\psi)\) gives
\(\mathtt{Ciuciura}_{4}(\varphi,\psi)\), and the same substitution pattern gives all higher Ciuciura axioms. Moreover, by Ciuciura's result \cite{Ciuciura2018}, none of the Ciuciura axiom schemata is derivable in $\mathbf{CLuN}$. Since $\mathbf{mbC}^\omega = \mathbf{CLuN}$, it follows that $\mathbf{mbC}^\omega$ is the weakest system in the hierarchy.

Proposition~\ref{hierarchy-inclusion} only shows that each system
derives the characteristic axioms of every weaker system. The hierarchy
is in fact strict: Ciuciura~\cite[Theorem~3.2.6]{Ciuciura2018} shows that \(\mathbf{mbC}^{n+1}\) is a~proper
subsystem of \(\mathbf{mbC}^{n}\) for every \(n\geq0\), so the diagram in
Figure~\ref{fig:mbc-hierarchy} depicts strict inclusions between
consecutive levels.

Semantically, the system $\mathbf{mbC}^{1}$ was studied by Omori \cite{Omori2019}, who developed a three-valued non-deterministic semantics for it. Before turning to our own semantic construction, which extends Omori's result, we first recall the semantics introduced by Ciuciura for the systems $\mathbf{mbC}^{n}$.

\begin{definition}[Ciuciura semantics for $\mathbf{mbC}^{n}$]
\label{def:ciuciura-semantics}
Fix \(n\geq 1\) and put \(k=n+1\). A \emph{\(\mathbf{mbC}^{n}\)-valuation} is a~function $v: \Fm \to \{0,1\}$, such that for all \(\varphi,\psi\in\Fm\):

\begin{enumerate}[label=(\roman*),leftmargin=2em]

\item \(v(\neg\varphi)=0 \Longrightarrow v(\varphi)=1\).

\item \(v(\n^{k}\varphi)=1 \Longrightarrow v(\varphi)=0
\text{ or } v(\n\varphi)=0 \text{ or } \cdots \text{ or }
v(\n^{k-1}\varphi)=0\).

\item \(v(\varphi\vee\psi)=1 \Longleftrightarrow
v(\varphi)=1 \text{ or } v(\psi)=1\).

\item \(v(\varphi\et\psi)=1 \Longleftrightarrow
v(\varphi)=1 \text{ and } v(\psi)=1\).

\item \(v(\varphi\jt\psi)=1 \Longleftrightarrow
v(\varphi)=0 \text{ or } v(\psi)=1\).

\end{enumerate}
\end{definition}

Validity and semantic consequence in $\mathbf{mbC}^{n}$ are defined in the usual way. A formula \(\varphi\) is \emph{valid}, written \(\models_{\mathbf{n}}\varphi\), iff $v(\varphi)=1$ for every \(\mathbf{mbC}^{n}\)-valuation \(v\).  We also define the consequence relation by saying \(\varphi\) is a \emph{semantic consequence} of a set \(\Gamma\), written \(\Gamma\models_{\mathbf{n}}\varphi\), iff there is no \(\mathbf{mbC}^{n}\)-valuation \(v\) such that $v(\psi)=1$ for all  $\psi\in\Gamma$, while $v(\varphi)=0$.

Clearly, the truth-conditions resemble the standard Boolean conditions for classical propositional logic. The only exception is negation. The semantics of negation is specified by two clauses. The first is a weakening of the classical truth condition, while the second regulates the behaviour of the $k$-th iteration of negation by requiring that for some $j<k$, the $j$-th iteration of the negation is false. Ciuciura proved that this semantics is sound and complete with respect to the axiomatization of $\mathbf{mbC}^{n}$.

\begin{theorem}[Soundness and completeness \cite{Ciuciura2014}]
    Let \(\Gamma \subseteq \Fm\) and \(\varphi \in \Fm\).
    \[
    \Gamma \vdashn \varphi \iff \Gamma \models_{\mathbf{n}}\varphi.
    \]
\end{theorem}

The following theorem motivates the move to non-deterministic semantics. It shows that for any level in the hierarchy $n\geq 1$ there is no finite deterministic matrix semantics that is sound and complete.

\begin{theorem}
For every \(n\geq 1\), the consequence relation of \(\mathbf{mbC}^{n}\) is not characterized by any finite deterministic matrix
\[
\mathcal M=\langle V,D,\neg^{\mathcal M},\wedge^{\mathcal M},\vee^{\mathcal M},\to^{\mathcal M}\rangle.
\]
\end{theorem}

\begin{proof}
The proof comes in two parts. In the first part, we define a formula and show that any finitely-many valued deterministic matrix characterizing $\mathbf{mbC}^{n}$ has to validate it. The second part shows that the formula itself is not valid in Cuciura's semantics and together with the fact that the semantics is complete with respect to $\mathbf{mbC}^{n}$ we arrive at a contradiction. Thus there is no such finite semantics.\footnote{This is a variant of the proof used in \cite{pawlowskifergusongertler2025} as well as a variant of Theorem 24 from \cite{Avron2011}.}

Fix $n\geq 1$ and suppose, towards a contradiction, that there exists a finite deterministic matrix \(\mathcal M\) characterizing \(\mathbf{mbC}^{n}\). Since \(V\) is finite, there are only finitely many unary functions on \(V\). Therefore, by a variation of the pigeonhole principle, there are $i,j \geq 0$ with $i<j$ such that
$(\neg^{\mathcal M})^{i}=(\neg^{\mathcal M})^{j}$. Here $(\neg^{\mathcal M})^{k}$ denotes the $k$-th iteration  (composition) of the negation operation with itself. So, there is a loop at some point since only finitely many iterates of a unary operation on a finite set can be distinct.

Let $p \in \Prop$ be an arbitrary propositional variable and define \(\theta_{ij}:=\neg^{i}p\to\neg^{j}p\). For every \(\mathcal M\)-valuation \(v: \Fm \to V\), we have \(v(\neg^{i}p)=v(\neg^{j}p)\) by the construction and the fact that the matrix only has finitely-many values. Observe that, since \(\mathbf{mbC}^{n}\) proves every instance of \(\varphi\to \varphi\) and $\mathcal{M}$ characterizes $\mathbf{mbC}^n$, the instantiation of the formula to formula \(p\to p\) has to be valid in \(\mathcal M\). This implies that \(p\) may receive any value \(a\in V\). Hence \(a\to^{\mathcal M}a\in D\) for every \(a\in V\). It follows that every \(\mathcal M\)-valuation designates \(\theta_{ij}\). Thus \(\models_{\mathcal M}\theta_{ij}\), and therefore \(\vdashn\theta_{ij}\).

We now show that \(\theta_{ij}\) is not valid in Ciuciura's semantics. Recall that in Ciuciura's semantics for \(\mathbf{mbC}^{n}\), where \(n\geq 1\) and \(k=n+1\), the negation clauses impose two restrictions on every sequence $v(\psi),v(\neg\psi),v(\neg^{2}\psi),\ldots$. First, clause (i) states that no two consecutive members of such a sequence may both be \(0\). Indeed, if
\(v(\neg^{m+1}\psi)=0\), then applying clause (i) to
\(\neg^{m}\psi\) gives \(v(\neg^{m}\psi)=1\).
Second, clause (ii) states that there can be no block of \(n+2\) consecutive \(1\)'s. To see this, suppose that there is such a sequence $v(\neg^{m}\psi)=v(\neg^{m+1}\psi)=\cdots
=v(\neg^{m+n+1}\psi)=1$. Clearly this block has length $n+2$. Hence we know that applying clause (ii) to \(\neg^{m+n+1}\psi\) implies that at least one of $\neg^{m}\psi,\neg^{m+1}\psi,\ldots,\neg^{m+n}\psi$ must have value \(0\), contradicting the assumption. So, in order to show that the formula in question is not valid, we need to  construct a sequence of values $b_0,b_1,b_2,\ldots\in\{0,1\}$ such that \(b_i=1\), \(b_j=0\), with no two consecutive \(0\)'s and no three consecutive \(1\)'s. Since \(n\geq 1\), it suffices to forbid three consecutive \(1\)'s, as it automatically excludes blocks of \(n+2\) consecutive \(1\)'s. We begin by setting \(d=j-i\). Recall that \(i<j\), so \(d\geq 1\). We first define the sequence from  position \(i\) onwards. Now, there are two cases. If \(d\) is odd, we define the sequence the following way:
\[
b_i,b_{i+1},b_{i+2},\ldots
=
1,0,1,0,\ldots .
\]
Then \(b_i=1\), and since \(d\) is odd, \(b_j=b_{i+d}=0\). Otherwise, if \(d\) is even:
\[
b_i,b_{i+1},b_{i+2},b_{i+3},\ldots
=
1,1,0,1,0,\ldots.
\]
Then again \(b_i=1\), and since \(d\) is even and positive, \(b_j=b_{i+d}=0\).
Extend the sequence to the left of \(i\) by alternating values, i.e., $b_{i-1}=0,\quad b_{i-2}=1,\quad b_{i-3}=0,\quad b_{i-4}=1$, and so on. In both cases the resulting sequence has no two consecutive \(0\)'s and no three consecutive \(1\)'s. Finally, build the valuation based on the sequence. We take  $v(\neg^{m}p)=b_m$ for $m\geq0$. In particular, $v(\neg^{i}p)=b_i=1$ and $v(\neg^{j}p)=b_j=0$. It remains to check that this partial assignment can be extended to a total \(\mathbf{mbC}^{n}\)-valuation. This is immediate, since the values of all iterated negations of $p$ have already been fixed, and the remaining formulas can be evaluated recursively using the truth conditions for the positive connectives. For this valuation $v(\neg^{i}p)=1$ and $v(\neg^{j}p)=0$, which yields $v(\neg^{i}p\to\neg^{j}p)=0$. Therefore \(\theta_{ij}\) is not valid in Ciuciura's semantics. This contradicts the assumption that $\mathcal{M}$ characterizes $\mathbf{mbC}^n$.
\end{proof}

In the next section, we begin by recalling Omori's results and then show how non-deterministic semantics can be extended to provide a natural characterization of the entire hierarchy.

\section{Non-deterministic semantics for the hierarchy}
We begin by recalling the non-deterministic semantic proposed by Omori in \cite{Omori2019} for the first level of the hierarchy. It is a three-valued semantics with the set of truth-values \(V=\{f,t,b\}\) and the set of designated values \(D=\{t,b\}\). For the sake of uniformity, we use a slightly different notation, identifying the values \(\mathtt{0}=f,\mathtt{1}=t, \mathtt{2}=b\). The connectives are interpreted via the following tables 
\[
\begin{minipage}{0.18\textwidth}
\centering
\[
\begin{array}{c|c}
\varphi & \nt \varphi\\
\hline
\mathtt{0} & \{\mathtt{1},\mathtt{2}\}\\
\mathtt{1} & \{\mathtt{0}\}\\
\mathtt{2} & \{\mathtt{1}\}
\end{array}
\]
\end{minipage}
\hfill
\begin{minipage}{0.25\textwidth}
\centering
\[
\begin{array}{c|ccc}
\wtwedge & \mathtt{0} & \mathtt{1} & \mathtt{2}\\
\hline
\mathtt{0} & \{\mathtt{0}\} & \{\mathtt{0}\} & \{\mathtt{0}\}\\
\mathtt{1} & \{\mathtt{0}\} & D & D\\
\mathtt{2} & \{\mathtt{0}\} & D & D
\end{array}
\]
\end{minipage}
\hfill
\begin{minipage}{0.25\textwidth}
\centering
\[
\begin{array}{c|ccc}
\wtvee & \mathtt{0} & \mathtt{1} & \mathtt{2}\\
\hline
\mathtt{0} & \{\mathtt{0}\} & D & D\\
\mathtt{1} & D & D & D\\
\mathtt{2} & D & D & D
\end{array}
\]
\end{minipage}
\hfill
\begin{minipage}{0.25\textwidth}
\centering
\[
\begin{array}{c|ccc}
\wtto & \mathtt{0} & \mathtt{1} & \mathtt{2}\\
\hline
\mathtt{0} & D & D & D\\
\mathtt{1} & \{\mathtt{0}\} & D & D\\
\mathtt{2} & \{\mathtt{0}\} & D & D
\end{array}
\]
\end{minipage}
\]

The construction for $\mathbf{mbC}^{n}$ is uniform. The positive connectives retain their classical behaviour at the level of designatedness. Whenever the corresponding classical connective returns \(1\), the non-deterministic operation returns the set of designated values; whenever it returns \(0\), it returns \(\{\mathtt{0}\}\). The only genuinely non-classical behaviour is exhibited by negation. We present the construction only for $n \geq 1$, since the cases of $\mathbf{CLuN}$ and $\mathbf{CPL}$ are well known and arise as degenerate cases of the same framework.

\begin{definition}[The $\mathbf{mbC}^{n}$-matrix]
Let \(n\geq 1\). The $\mathbf{mbC}^{n}$-matrix is the Nmatrix \(\Mn=\langle V_n,D_n,O_n\rangle\), where \(V_n\) is the set of truth-values, \(D_n\subseteq V_n\) is the set of designated values, and \(O_n=\{\nt,\wtwedge,\wtvee,\wtto\}\) is the set of non-deterministic operations for the connectives. These are defined as follows:
\begin{enumerate}[label=(\roman*),leftmargin=2em]
\item \(V_n=\{\mathtt{0},\mathtt{1},\ldots,\mathtt{n+1}\}\).

\item \(D_n=\{\mathtt{1},\ldots,\mathtt{n+1}\}\).

\item The interpretations of the connectives are given by

\[
\nt a=
\begin{cases}
D_n, & \text{if } a=\mathtt{0},\\
\{\mathtt{0}\}, & \text{if } a=\mathtt{1},\\
\{\mathtt{k-1}\}, & \text{if } a=\mathtt{k}\text{ for some }2\leq k\leq n+1.
\end{cases}
\]

and  for each binary connective \(\circledast\in\{\wedge,\vee,\to\}\), the corresponding interpetation is

\[
a\,\interp{\circledast}\,b=
\begin{cases}
D_n, & \text{if } \circledast=\wedge \text{ and } a,b\in D_n,\\
D_n, & \text{if } \circledast=\vee \text{ and } (a\in D_n \text{ or } b\in D_n),\\
D_n, & \text{if } \circledast=\to \text{ and } (a=\mathtt{0} \text{ or } b\in D_n),\\
\{\mathtt{0}\}, & \text{otherwise}.
\end{cases}
\]

\end{enumerate}
\end{definition}

The clauses above have a simple reading. The value \(\mathtt{0}\) is the only undesignated value. The values \(\mathtt{1},\ldots,\mathtt{n+1}\) are all designated. The idea is that they record how many applications of negation may still be made before the value reaches \(\mathtt{0}\). Thus, if a formula receives value \(\mathtt{k}\), with \(1\leq k\leq n+1\), then its negation receives \(\mathtt{k-1}\). Only when \(k=1\), the negation yields the undesignated value \(\mathtt{0}\). By contrast, when a formula receives the value \(\mathtt{0}\), its negation may receive any designated value. This is the only point at which the operation is genuinely non-deterministic. The positive connectives behave classically with respect to designatedness as in Ciuciura's original semantics.

The parameter \(n\) fixes the number of designated levels available above \(\mathtt{0}\). Increasing \(n\) therefore allows longer sequences of iterated negations to remain designated before an undesignated value must appear. For \(n=1\), the values are precisely \(\mathtt{0},\mathtt{1},\mathtt{2}\), and the definition gives Omori's three-valued semantics for $\mathbf{mbC}^{1}$ up to the renaming \(\mathtt{0}=f\), \(\mathtt{1}=t\), and \(\mathtt{2}=b\).

\begin{lemma}[Values of iterated negations]\label{lem:rank-chain} Let \(v\) be a  \(\Mn\)-valuation. For every formula \(\varphi\), the following holds: \[ v(\n^{j}\varphi)= \begin{cases} \mathtt{0}, & \text{if } v(\varphi)=\mathtt{0}\text{ and }j=0,\\ \mathtt{k-j}, & \text{if } v(\varphi)=\mathtt{k}\text{ for some }1\leq k\leq n+1 \text{ and }0\leq j  <k,\\ \mathtt{0}, & \text{if } v(\varphi)=\mathtt{k}\text{ for some }1\leq k\leq n+1 \text{ and }j=k. \end{cases} \] Consequently, at least one formula among \(\varphi,\n\varphi,\ldots,\n^{n+1}\varphi\) receives value \(\mathtt{0}\). \end{lemma}

\begin{proof}
If \(v(\varphi)=\mathtt{0}\), then the first condition is trivially satisfied if $j=0$ and furthermore  the first formula in the list \(\varphi,\n\varphi,\ldots,\n^{n+1}\varphi\) already receives value
\(\mathtt{0}\). Let us now assume  that \(v(\varphi)=\mathtt{k}\), where
\(1\leq k\leq n+1\). We will prove  by induction on \(j\), \(0\leq j<k\), that
\(v(\n^{j}\varphi)=\mathtt{k-j}\). If \(j=0\), then by the assumption  \(v(\varphi)=\mathtt{k}\). Now suppose that \(0\leq j<k-1\) and that
\(v(\n^{j}\varphi)=\mathtt{k-j}\). Since \(j<k-1\), we have
\(k-j>1\). Hence, by the definition of negation in \(\Mn\), $\nt\mathtt{k-j}=\{\mathtt{k-j-1}\}.$ By the fact that \(v\) is a \(\Mn\)-valuation, it follows that $v(\n^{j+1}\varphi)=\mathtt{k-j-1}$. This proves the induction step, so \(v(\n^{j}\varphi)=\mathtt{k-j}\) for every \(0\leq j<k\).
In particular, taking \(j=k-1\), we get
\(v(\n^{k-1}\varphi)=\mathtt{1}\). Since
\(\nt\mathtt{1}=\{\mathtt{0}\}\), it follows that
\(v(\n^{k}\varphi)=\mathtt{0}\). As \(k\leq n+1\), the formula
\(\n^{k}\varphi\) occurs among
\(\varphi,\n\varphi,\ldots,\n^{n+1}\varphi\). Hence at least one formula
in this list receives value \(\mathtt{0}\).
\end{proof}

The next step is to prove the soundness of our semantics.

\begin{theorem}[Soundness]\label{thm:soundness}
For all \(\Gamma\cup\set{\varphi}\subseteq\Fm, \mathbf{n}\geq 1\),
\[
\Gamma\vdashn \varphi\quad\Longrightarrow\quad \Gamma\modelsM \varphi.
\]
\end{theorem}
\begin{proof}
Fix \(n\geq 1\), and let \(v\) be an arbitrary \(\Mn\)-valuation. The positive fragment \((\mathrm A_{1}-\mathrm A_{9})\) is relatively straightforward, as $\wedge,\vee,\to$ behave classically. For \((\mathrm A_{10})\), namely \(\psi\vee\n\psi\), if \(v(\psi)\in D_n\), then the disjunction is designated. If \(v(\psi)=\mathtt{0}\), then \(v(\n\psi)\in D_n\) by the negation table, and hence the disjunction is designated.

For the Ciuciura's axiom, consider an instance \(\psi\to(\n\psi\to(\n^{2}\psi\to\cdots(\n^{n+1}\psi\to\chi)\cdots))\). By Lemma~\ref{lem:rank-chain}, at least one of \(\psi,\n\psi,\ldots,\n^{n+1}\psi\) has value \(\mathtt{0}\). Let \(\n^{r}\psi\) be the first such formula in the displayed chain. The subimplication beginning with antecedent \(\n^{r}\psi\) is designated, since an implication with antecedent \(\mathtt{0}\) is designated. Moving outwards through the nested implication, every preceding implication has a designated consequent, and is therefore designated. Thus every instance of \(\mathtt{Ciuciura}_{n+1}\) is valid.

Finally, modus ponens preserves designatedness. Suppose \(v(\psi)\in D_n\) and \(v(\psi\to\chi)\in D_n\). If \(v(\chi)=\mathtt{0}\), then, since \(v(\psi)\in D_n\), the implication table gives \(v(\psi\to\chi)=\mathtt{0}\), contrary to \(v(\psi\to\chi)\in D_n\). Hence \(v(\chi)\in D_n\). Therefore MP preserves designatedness.

Since every axiom is valid and the only inference rule preserves designatedness, induction on derivations yields the claim: if \(\Gamma\vdashn\varphi\), then every \(\Mn\)-valuation that designates all formulas in \(\Gamma\) also designates \(\varphi\). Hence \(\Gamma\modelsM\varphi\).
\end{proof}

We home in on the famously, or infamously, more complicated direction, namely the completeness. The proof is quite standard. We will use the canonical-function construction that is based on the notion of a saturated set.

\begin{definition}[Saturated sets]\label{def:saturated}
Let \(\varphi\in\Fm\). A set \(\Pi\subseteq\Fm\) is \(\varphi\)-saturated, relative to $\mathbf{mbC}^{n}$, iff
\begin{enumerate}[label=(\roman*),leftmargin=2em]
\item \(\Pi\nvdashn \varphi\), and
\item for every \(\psi\in\Fm\), if \(\psi\notin\Pi\), then \(\Pi\cup\set{\psi}\vdashn\varphi\).
\end{enumerate}
\end{definition}

\begin{lemma}[Saturated extension]\label{lem:saturated-extension}
If \(\Gamma\nvdashn\varphi\), then there is a \(\varphi\)-saturated set \(\Pi\) such that \(\Gamma\subseteq\Pi\).
\end{lemma}

\begin{proof}
Let \(\mathcal S\) be the family of sets \(\Delta\) with \(\Gamma\subseteq\Delta\subseteq\Fm\)
and \(\Delta\nvdashn\varphi\), ordered by inclusion. \(\mathcal S\) is non-empty, since
\(\Gamma\in\mathcal S\) by hypothesis. The union of any chain in \(\mathcal S\) stays in
\(\mathcal S\): it clearly contains \(\Gamma\), and a derivation of \(\varphi\) from the union
would use only finitely many premises, all already contained in some single member of the
chain, contradicting \(\Delta\nvdashn\varphi\) for that member. By Zorn's lemma, \(\mathcal S\)
has a maximal element \(\Pi\), and \(\Pi\nvdashn\varphi\) by construction. If \(\psi\notin\Pi\),
then \(\Pi\cup\set\psi\notin\mathcal S\) by maximality of \(\Pi\), and since
\(\Gamma\subseteq\Pi\cup\set\psi\), this forces \(\Pi\cup\set\psi\vdashn\varphi\). Hence \(\Pi\)
is \(\varphi\)-saturated.
\end{proof}
\begin{lemma}[Facts for saturated sets]\label{lem:sat-connectives}
Let \(\Pi\) be \(\varphi\)-saturated. Then, for all formulas \(\psi,\chi\), the following hold:
\begin{enumerate}[label=\textup{(\roman*)},leftmargin=2em]
    \item \(\Pi\vdashn \psi\wedge\chi\) iff \(\Pi\vdashn \psi\) and \(\Pi\vdashn \chi\).

    \item \(\Pi\vdashn \psi\vee\chi\) iff \(\Pi\vdashn \psi\) or \(\Pi\vdashn \chi\).

    \item \(\Pi\vdashn \psi\to\chi\) iff \(\Pi\nvdashn \psi\) or \(\Pi\vdashn \chi\).

    \item \(\Pi\vdashn \psi\) or \(\Pi\vdashn\n\psi\).

    \item Not all of \(\psi,\n\psi,\ldots,\n^{n+1}\psi\) are derivable from \(\Pi\).
\end{enumerate}
\end{lemma}

\begin{proof}
Conjunction (\textup{i}) follows directly from applications of the axioms \((\mathrm{A}_3-\mathrm{A}_5)\) together with \emph{modus ponens}. For disjunction (\textup{ii}), the right-to-left direction follows from \((\mathrm{A}_6)\) and \((\mathrm{A}_7)\) by \emph{modus ponens}. For the  left-to-right direction, suppose  for contradiction that \(\Pi\vdashn\psi\vee\chi\) while \(\Pi\nvdashn\psi\) and \(\Pi\nvdashn\chi\). Then \(\psi,\chi\notin\Pi\), so saturation gives \(\Pi\cup\set{\psi}\vdashn\varphi\) and \(\Pi\cup\set{\chi}\vdashn\varphi\). By the deduction theorem\footnote{ For the sake of clarity, notice that the deduction theorem holds for $\mathbf{mbC}^{n}$ \cite{Ciuciura2014}, since the only rule is \emph{modus ponens}, and the positive basis contains the standard classical propositional implication.} we get \(\Pi\vdashn\psi\to\varphi\) and \(\Pi\vdashn\chi\to\varphi\), and \((\mathrm{A}_8)\) with \(\Pi\vdashn\psi\vee\chi\) yields \(\Pi\vdashn\varphi\), contradicting the saturation of $\Pi$. For implication (\textup{iii}), the left-to-right direction follows immediately by \emph{modus ponens}. Conversely, notice that if \(\Pi\vdashn\chi\), then \((\mathrm{A}_1)\) and \emph{modus ponens} give \(\Pi\vdashn\psi\to\chi\). On the other hand, if  \(\Pi\nvdashn\psi\), then \((\mathrm{A}_9)\) yields \(\Pi\vdashn\psi\vee(\psi\to\chi)\), whence the disjunction clause gives \(\Pi\vdashn\psi\to\chi\).   For \textup{(iv)}, let \(\psi\) be arbitrary. Since
\((\mathrm A_{10})\) is an axiom schema, we have
\(\vdashn \psi\vee\n\psi\), and hence also
\(\Pi\vdashn \psi\vee\n\psi\). By \textup{(ii)}, applied to the
disjunction \(\psi\vee\n\psi\), it follows that
\(\Pi\vdashn\psi\) or \(\Pi\vdashn\n\psi\). This proves \textup{(iv)}. Finally, were \(\psi,\n\psi,\ldots,\n^{n+1}\psi\) all derivable, \(\mathtt{Ciuciura}_{n+1}(\psi,\varphi)\) and repeated application of \emph{modus ponens} would give \(\Pi\vdashn\varphi\), again contradicting the saturation of $\Pi$.
\end{proof}

\begin{definition}[Canonical function]\label{def:canonical-function}
Let \(\Pi\) be \(\varphi\)-saturated. Define \(v_\Pi:\Fm\to V_n\) by the following case distinction:
\[
v_\Pi(\psi)=
\begin{cases}
\mathtt{0}, & \text{if } \Pi\nvdashn \psi,\\[.8ex]
\mathtt{k}, &
\begin{array}[t]{l}
\text{if } \Pi\vdashn \psi,\text{ where }k\text{ is the least number such that}\\
1\leq k\leq n+1\text{ and }\Pi\nvdashn \n^{k}\psi.
\end{array}
\end{cases}
\]
\end{definition}

The function $v_\Pi$ is well defined. If \(\Pi\nvdashn\psi\), the first case applies. If \(\Pi\vdashn\psi\), by Lemma~\ref{lem:sat-connectives}(v), not all of \(\psi,\n\psi,\ldots,\n^{n+1}\psi\) are derivable from $\Pi$. Hence there exists a least \(k\), with \(1\leq k\leq n+1\), such that $\Pi \nvdashn \neg^k \psi$.
\begin{lemma}[Canonical function lemma]\label{lem:canonical}
For every formula \(\psi\),
\[
v_\Pi(\psi)\in D_n\quad\text{iff}\quad \Pi\vdashn \psi.
\]
Moreover, \(v_\Pi\) is a  \(\Mn\)-valuation.
\end{lemma}

\begin{proof}
The equivalence is immediate from Definition~\ref{def:canonical-function}. To show that the canonical function $v_\Pi$ is a~valuation, we rely on the equivalence above and the conditions of the saturated sets proved in Lemma~\ref{lem:sat-connectives}. We observe that:

\begin{enumerate}
\item \(v_\Pi(\psi\wedge\chi)\) is designated exactly when \(v_\Pi(\psi)\) and \(v_\Pi(\chi)\) both are, as required by the interpretation of \(\wtwedge\).
\item   The cases for \(\vee\) and \(\to\) are analogous, using the corresponding clauses of Lemma~\ref{lem:sat-connectives}. 
\item  For negation: if \(v_\Pi(\psi)=\mathtt{0}\), then \(\Pi\nvdashn\psi\), so \(\Pi\vdashn\n\psi\) and \(v_\Pi(\n\psi)\in D_n=\nt\mathtt{0}\). If \(v_\Pi(\psi)=\mathtt{1}\), then \(\Pi\vdashn\psi\) and \(\Pi\nvdashn\n\psi\), so \(v_\Pi(\n\psi)=\mathtt{0}\in\nt\mathtt{1}\). If \(v_\Pi(\psi)=\mathtt{k}\) with \(k>1\), then \(\Pi\) derives \(\psi,\n\psi,\ldots,\n^{k-1}\psi\) but not \(\n^{k}\psi\), so the derivable chain from \(\n\psi\) has length \(k-1\) and \(v_\Pi(\n\psi)=\mathtt{k-1}\in\nt\mathtt{k}\).
\end{enumerate}
We conclude that \(v_\Pi\) is a \(\Mn\)-valuation.
\end{proof}

\begin{theorem}[Strong completeness]\label{thm:completeness}
For all \(\Gamma\cup\set{\varphi}\subseteq\Fm\),
\[
\Gamma\modelsM \varphi\quad\Longrightarrow\quad \Gamma\vdashn \varphi.
\]
\end{theorem}

\begin{proof}
By contraposition, suppose \(\Gamma\nvdashn\varphi\). Lemma~\ref{lem:saturated-extension} supplies a \(\varphi\)-saturated \(\Pi\supseteq\Gamma\), and Lemma~\ref{lem:canonical} makes \(v_\Pi\) a \(\Mn\)-valuation. Each \(\gamma\in\Gamma\) lies in \(\Pi\), hence \(\Pi\vdashn\gamma\) and \(v_\Pi(\gamma)\in D_n\); since \(\Pi\nvdashn\varphi\), Definition~\ref{def:canonical-function} gives \(v_\Pi(\varphi)=\mathtt{0}\). Thus \(\Gamma\nmodelsM\varphi\).
\end{proof}

\section{Conclusions}

In this paper, we introduced a uniform non-deterministic semantics for the hierarchy of systems $\mathbf{mbC}^{n}$ introduced by Ciuciura. The construction generalizes Omori’s three-valued semantics for $\mathbf{mbC}^{1}$ to all levels of the hierarchy. This is done using $n+2$-values to characterize $\mathsf{mbC}^n$. We also proved that the consequence relations of the systems $\mathbf{mbC}^{n}$ cannot be characterized by finite deterministic matrices. This further motivates the use of Nmatrices in the study of the hierarchy. It remains open and superbly fruitful whether similar uniform Nmatrix semantics can be obtained for other families of paraconsistent logics.

\paragraph{Use of AI and Funding}
Generative AI tools were used to assist with linguistic editing, LaTeX troubleshooting, and exploratory brainstorming about possible proof strategies and conceptual connections between the systems. All final formulations, mathematical arguments, examples, references, and claims were independently checked by the authors. 
Pawel Pawlowski has been supported by the FWO senior-postdoctoral grant 1255724N.
\bibliographystyle{eptcs}
\bibliography{haptism}
\end{document}